\documentclass[twocolumn,runningheads]{svjour2}
\smartqed  
\usepackage[active]{srcltx}
\usepackage{graphicx}
\usepackage{mathptmx}   
\usepackage{amsmath}
\usepackage{amsfonts}
\usepackage{amssymb}
\usepackage{latexsym}
\usepackage[authoryear]{natbib}
\usepackage{journals}

\journalname{Astrophysics and Space Science (CoRoT/ESTA Volume)}
\newcommand{\logg}{\mbox{${\log g}$}}

\newcommand{\msol}{\mbox{${M}_{\odot}$}}
\newcommand{\lsol}{\mbox{${\mathrm L}_{\odot}$}}

\newcommand{\corot}{{\small CoRoT}}

\newcommand{\esta}{{\small ESTA}}

\newcommand{\cesam}{{\small CESAM}}
\newcommand{\cles}{{\small CL\'ES}}

\newcommand{\postMS}{{\small Post-MS}}
\newcommand{\PMS}{{\small PMS}}
\newcommand{\MS}{{\small MS}}

\newcommand{\HR}{{\small HR}}
\newcommand{\HRD}{{\small HRD}}

\newcommand{\losc}{{\small LOSC}}
\newcommand{\thisapss}{Astrophys. Space Sci. (CoRoT/ESTA Volume)~}

\begin{document}

\title{Grids of Stellar Models and Frequencies with CL\'ES + LOSC 
}
\subtitle{}

\titlerunning{Grids:\cles+\losc}        

\author{  Josefina Montalb\'an        \and
          Andrea Miglio  \and
	  Arlette Noels  \and
          Richard Scuflaire 
}


\institute{J.~Montalb\'an \and A.~Miglio \and A.~Noels \and R.~Scuflaire
            \at 
            Institut d'Astrophysique et Geophysique, Universit\'e de Li\`ege,
            all\'ee du 6 Ao\^ut 17, B-4000 Li\`ege, Belgium\\
           \email{J.Montalban, A.Miglio, Arlette.Noels, R.Scuflaire@ulg.ac.be}}

\date{Received: date / Accepted: date}

\maketitle
\begin{abstract}

We present a grid of stellar models, obtained with the \cles\
evolutionary
code, following the specification of  \esta-Task1, and the
corresponfing seismic properties,
computed with the \losc\ code. We provide a complete description of
the corresponding
files that will be available on the \esta\ web-pages.

\end{abstract}


\section{Introduction}\label{sec:intro}

The preparation of the \corot\ mission brought up the need of a
reference grid of stellar models in order to locate the possible \corot\ targets
in the Hertzsprung-Russell  (\HR) diagram, and to allow a first interpretation
of the forthcoming \corot-data. 
For this purpose,several grids of stellar models were computed according to the \esta-Task1
specifications.
In this paper we present the grids of stellar models we computed with the code 
\cles\ \citep{rs1-apss}, as well as the seismic properties  we derived by using 
the adiabatic oscillation code \losc\ \citep{rs2-apss} for models on the main sequence
(\MS) or close to it.

In sections~\ref{fisica} and \ref{starparam}  we describe, respectively, 
the input physics and the set of stellar parameters  we used. 
The evolutionary tracks, stellar models and oscillation frequency data can be found on the
\esta\ Web site. The description of all this
material is given in Sect.\ref{data} for the evolution and structure models, and in 
Sect.\ref{freqs} for the seismic properties of the stellar models.
Finally, in Sect.~\ref{conclu} we summarize the type and amount of available files.

\section{Input physics}
\label{fisica}

\begin{figure*}[htbp]
\centering
{\includegraphics[scale=0.8]{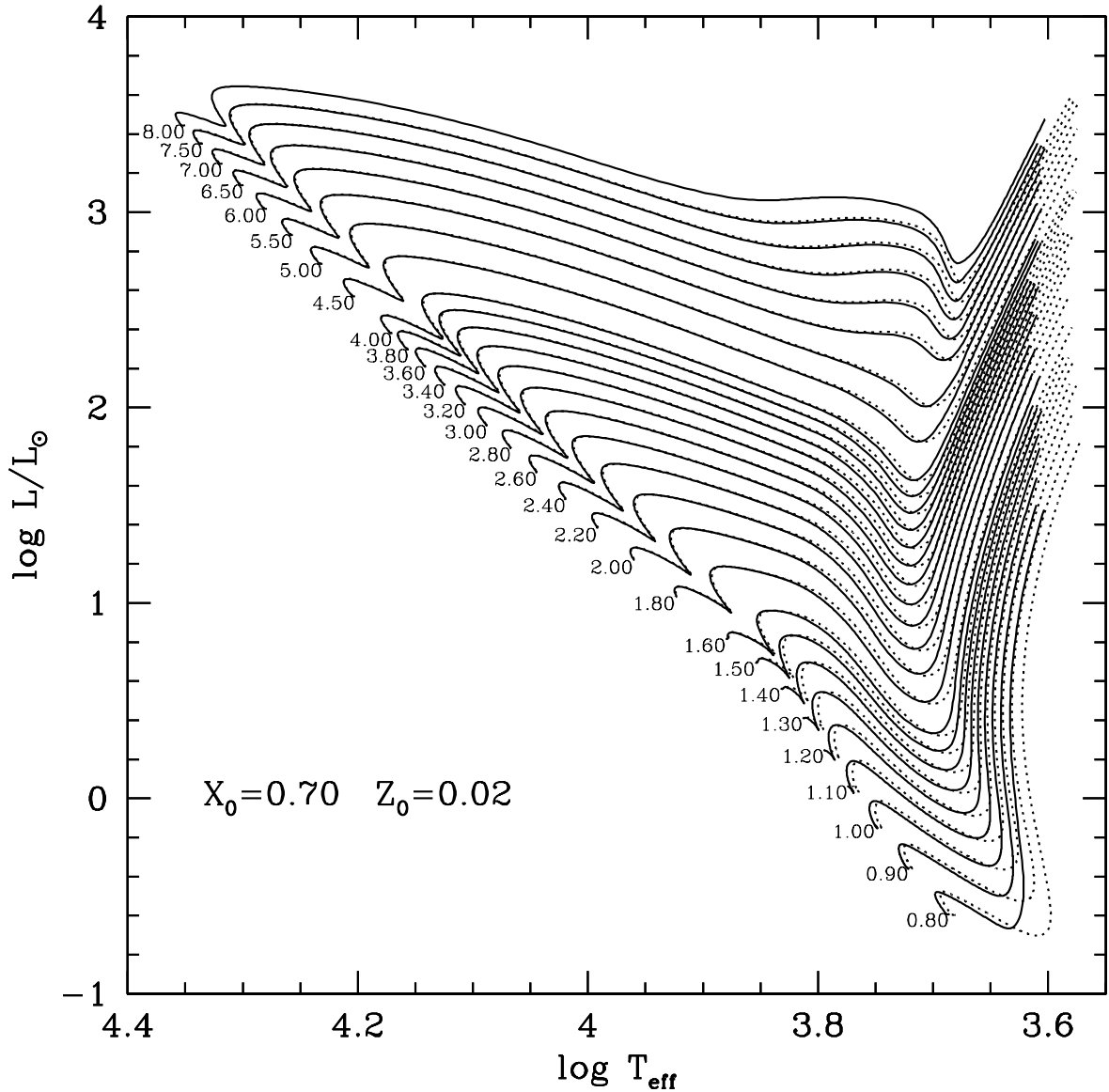}}
\caption{Theoretical \HR\ diagram for the evolution phase comprised 
 between the initial homogeneous model  and
that with a hydrogen mass fraction at the center  reduced by 1\% with respect to the initial
model (that is, $X_{\rm c}=0.693$). 
Evolutionary tracks are labeled by the corresponding stellar mass.
Solid lines correspond to grey models and dashed ones to  models with Kurucz boundary conditions.} 
\label{pms}
\end{figure*}

\begin{figure*}[htbp]
\centering
{\includegraphics[scale=0.8]{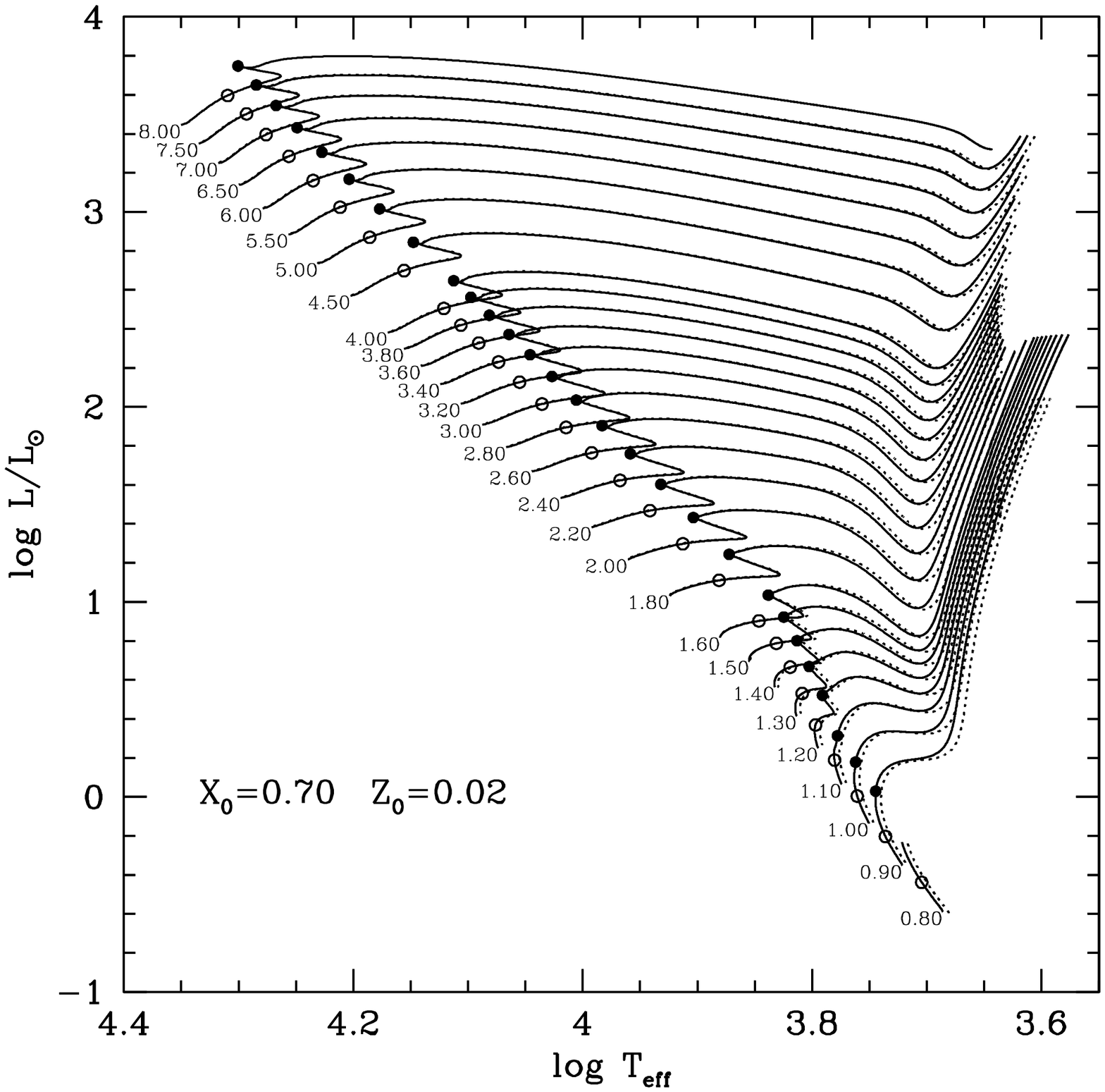}}
\caption{As Fig.~\ref{pms} for the evolutionary phase of main sequence and post-main sequence.
Open circles indicate the location of models that has burned, at the center, 
 half of the initial hydrogen. Solid circles correspond to models with only 0.1\% of hydrogen 
at the center.} 
\label{ms}
\end{figure*}

 For three out of four stellar model grids we adopted the
reference input physics, as well as the physical and astronomical constants
specified  for \esta\-Task~1 comparisons \citep[see also Sect.5 in][]{yl1-apss}.
We stress that these specifications do not always coincide with those of 
the standard version of \cles\ \citep[see][]{rs1-apss}.
\begin{itemize}
\item {\bf Equation of State (EoS)} --  OPAL~2001  equation of State 
\citep{2002ApJ...576.1064R} available in the OPAL Web site.
The tabulated values of $C_{\rm V}$ have been replaced with the values derived from the tabulated values of
$P$, $\Gamma_1$, $\chi_{\rm T}$, $\chi_{\rho}$. For each $\rho$, $T$, $X$ and $Z$ we interpolate in the tables
by using our own interpolation routine.

\item{\bf Opacities} -- OPAL96 \citep{ir96}. 
The opacity tables used in \cles\ have been calculated on-line for the standard
GN93 \citep{GN93}  metal mixture, and  using the smoothing  routine available in the 
OPAL Web site to build the final opacity tables. The low temperature opacity tables by 
\citet{af94} are also smoothly added \citep[see][]{rs1-apss}, and conductive opacities 
were not included in these computations.  As for  EoS, we interpolate in $\rho$, $T$, $X$ and $Z$ 
by using our own interpolation routines.

\item{\bf Nuclear network} -- We choose the same nuclear network as in
model comparison, that is, basic pp chain and CNO cycle reactions up to the $^{17}\rm O(p,\alpha)^{14}\rm N$. 
 We adopted the nuclear reaction rates from the analytical formulae provided by the NACRE
compilation \citep{1999NuPhA.656....3A}, including that one for the $^{14}{\rm N}(p,\gamma)^{15}{\rm O}$, 
and the weak screening factors from \citet{1954AuJPh...7..373S}.
The electronic density, taking part in the screening factor computation,  was estimated assuming full ionization 
of chemical elements.

\item{\bf Chemical composition} -- We adopted the metal distribution provided by the \citet{GN93} (thereafter GN93)
 solar mixture. Only the abundances of light elements (Li, Be and B) were modified to be consistent
with those used in \cesam. The isotopic ratios are those of \citet{AG89} except for
the $^2$H/$^1$H and  $^3$He/$^4$He  ratios for which we took those of \citet{GautierMorel}.

\item{\bf Convection} --  All the evolutionary tracks were computed  with the classical mixing length treatment
of convection by  \citet{1958ZA.....46..108B} and the formulation of \citet{1965ApJ...142..841H} 
for optically thin regions. The mixing length parameter $\alpha_{\rm MLT}$ was fixed at 1.6 for all the grids.
Note that the latter is the value chosen for ESTA comparisons, but it does not correspond to the
value derived from a solar calibration.

\item{\bf Overshooting.} All the models, regardless of their mass, were computed without convective core overshooting.

\item{\bf Atmosphere.} \esta--Task1 specification requires  the surface boundary conditions to be provided by 
 integration of a grey atmosphere following the Eddington's $T(\tau)$ law (grey models, thereafter).
 Three of the four stellar model grids were computed following the \esta\ specification, 
 while  for the  fourth one, temperature and density at $T=T_{\rm eff}$  were obtained  from  Kurucz's 
 atmosphere models \citep{kurucz98}.
\end{itemize}

\section{Stellar parameters}
\label{starparam}
We provide evolutionary tracks for masses from 0.8 to 8~\msol,
 with a mass step $\Delta M$=0.1, from 0.8 to 1.6~\msol,
$\Delta M$=0.2 up to 4.0~\msol, and $\Delta M$=0.5 from 4.0 to 8.0~\msol.

The initial hydrogen mass fraction is $X=0.70$ in all  grids. Three different values of
the  metal mass fraction $Z$ (0.02, 0.01 and 0.006)  are available for 
grids of grey models, while only $Z=0.02$ has been considered in the grid computed with  
Kurucz atmosphere boundary conditions.

\section{Data}

\label{data}
\subsection{Evolutionary tracks}

For each stellar parameter ($M,X,Z$) we follow the \PMS\ evolution from the Hayashi track, and
the calculation ends either when the temperature at the stellar center is high enough to burn He, or
when the age of the model is larger than 20~Gyr. We recall that the  computation of these models
does not include conductive opacities and that these ones  should be taken into account in modeling
low mass stars up to the helium ignition.
The grids of these evolutionary tracks are available from the ESTA Web 
site\footnote{http://www.astro.up.pt/corot/models} and 
consist of {\bf two}  \HRD-files, whose name contains the values of the stellar parameters
for which they were computed. For instance, ``m1.00Z0.02X0.70-\HRD1.txt'' and ``m1.00Z0.02X0.70-\HRD2.txt''
contain the evolutionary tracks for a 1.00~\msol\ star with initial 
chemical composition $X=0.70$ and $Z=0.02$.
The first four lines in \HRD1 and \HRD2-files  provide information about the stellar parameters and the 
input physics used to build the model. 
Since most of them have already been described in previous section, we only mention that  {\bf fg} 
and {\bf ft} indicate the factor by which the mesh and time step were multiplied.
Both, fg and ft were set equal to 0.5, that is, the number of  mesh points and the temporal 
steps were doubled with respect to the standard value used in \cles\ \citep{rs2-apss}.
The quantities for which the temporal evolution were tabulated  in the \HRD s files are the following:

\HRD1 files:
\begin {itemize}
\item {\bf Column 1} -- Stellar mass in units of \msol.
\item {\bf Column 2} -- Stellar luminosity ($\log (L/L_{\odot})$).
\item {\bf Column 3} -- Decimal logarithm of the effective temperature in K  ($\log T_{\rm eff}$).
\item {\bf Column 4} -- Stellar radius in units of solar radius ($R/R_{\odot}$).
\item {\bf Column 5} -- Stellar age in Myr.
\item {\bf Column 6} -- Hydrogen mass fraction at the center.
\item {\bf Column 7} -- \logg , with $g$, the  surface gravity .
\item {\bf Column 8} -- Number of convective-radiative boundaries.
\item {\bf Column 9} -- Codification of the of boundary type: ``2'' for boundary from convective to
 radiative region ;and ``1'' for boundaries from radiative  to convective one, going  
from the stellar center towards the surface.
\item {\bf Column 10-15} -- Relative radius of the six first convective-radiative boundaries
 (from the center outwards).
\end{itemize}

\HRD2 files:
\begin {itemize}
\item {\bf Column 1} -- Number of model.
\item {\bf Column 2} -- Stellar age in Myr.
\item {\bf Column 3} -- Stellar radius in solar radius ($R/R_{\odot}$).
\item {\bf Column 4} -- Stellar luminosity in units of solar luminosity ($L/$\lsol).
\item {\bf Column 5} -- Effective temperature in K.
\item {\bf Column 6} -- Temperature at the center in units of $10^7$~K.
\item {\bf Column 7} -- Central density $\rho_c$ in g.cm$^{-3}$.
\item {\bf Column 8} -- Hydrogen mass fraction at the center ($X_{\rm c}$).
\item {\bf Column 9} -- Central helium mass fraction ($Y_{\rm c}$).
\item {\bf Column 10} -- Number of convective regions.
\item {\bf Column 11} -- Relative mass  of the convective core ($m_{\rm cc}/M_{\ast}$).
\item {\bf Column 12} -- Relative radius of the base of the outer convective region ($R_{\rm ce}/R_{\ast}$).
\item {\bf Column 13} -- Mass of the helium core in \msol. 
The helium core is defined by the mass of the region with a hydrogen mass fraction smaller than $10^{-2}$.
\end{itemize}

\begin{figure}[htbp]
\centering
{\includegraphics[scale=0.4]{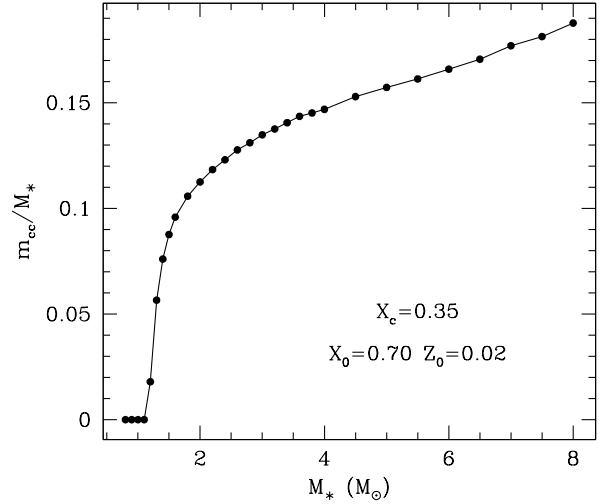}}
\caption{Relative mass of the convective core as a function of the stellar mass for models that
have burned the half of their initial H content at the center.} 
\label{mcc}
\end{figure}

\begin{figure}[htbp]
\centering
{\includegraphics[scale=0.4]{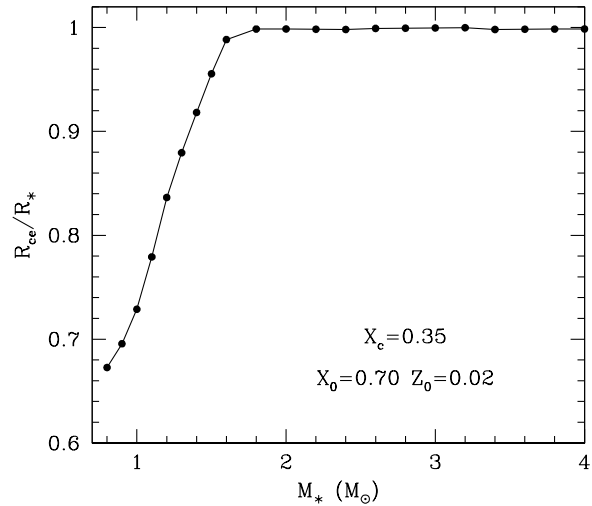}}
\caption{Relative radius of the outer  envelope bottom   as a function of the
stellar mass  for the same models as in Fig.~\ref{mcc}.} 
\label{rce}
\end{figure}

\begin{figure}[htbp]
\centering
{\includegraphics[scale=0.4]{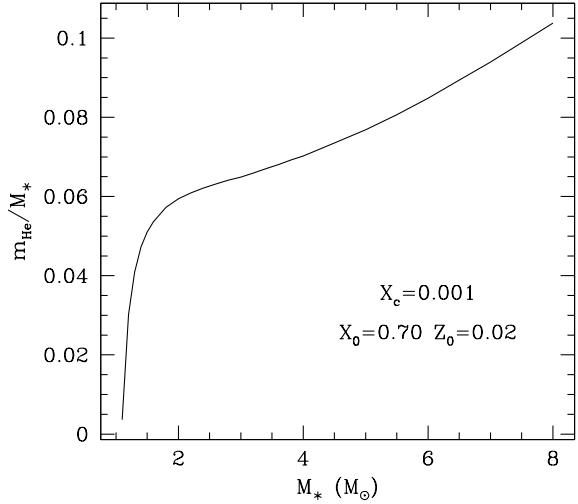}}
\caption{Relative mass of the helium core at the end of the \MS\ phase as a function of the 
stellar mass.} 
\label{mhe}
\end{figure}

The $\log T_{\rm eff}$--$\log L/L_{\odot}$ diagram for all the models with a metal mass fraction $Z=0.02$
have been plotted in Fig.~\ref{pms} for the \PMS\ evolution, and in Fig.~\ref{ms} 
for the \MS\ and \postMS\ evolutionary phases.
For each mass there is a pair of curves, one was computed with Eddington's law as boundary
conditions at the photosphere (solid lines) and the other used  Kurucz's atmosphere models 
(dashed lines). It is interesting to note that, as it concerns the theoretical \HR\ diagram, 
the  atmosphere type affects only models with effective temperature lower than 6300~K. 
The effects on the seismic properties will be  discussed in  Sect.~\ref{freqs}.

In Fig.~\ref{ms} we marked the model locations  at the middle
of their \MS\ (central hydrogen mass fraction $X_{\rm c}=0.35$) by open circles  and that of models
with a central hydrogen mass fraction $X_{\rm c} \simeq 0.001$ by  filled ones.
Defining the beginning of \MS\ by the model with a  central mass fraction of hydrogen
decreased by 1\% with respect to the initial one (that is $X_{\rm c}=0.693$ for the present computations), 
and its end  by that with $X_{\rm c} \simeq 0.001$, the \MS-lifetime varies from 
$5\times10^7$~yrs for
8~\msol\ to more than 13~Gyr for 0.9~\msol.  The number of computed stellar models that 
span the main sequence phase of evolution is of the order of 160  for each ($M,X,Z$). 

The dependence on the stellar mass of the convective core mass,  for models in the middle of 
\MS\, is shown in Fig.~\ref{mcc}, and that of the convective envelope bottom   in Fig.~\ref{rce}.
Finally, in Fig.~\ref{mhe} we show the mass of the He core at the end of the \MS\ as a function of
the stellar mass.

\subsection{Stellar models}

In addition to the evolutionary tracks,  a selection of  internal structure models for each $(M,X,Z)$ are also available :
\begin{itemize}
\item {\bf ZAMS} model - defined as the model for which the ratio of gravitational  to nuclear luminosity is lower  than 2\%.
\item {\bf 1MS} model - as the model which has burned 1\% of H  ($X_{\rm c} \simeq 0.693$).
\item {\bf MS} model   - as the model with  half of the initial H at the center ($X_{\rm c} \simeq 0.35$)
\item {\bf TAMS} model - as the model  with $X_{\rm c} \simeq 0.01$
\item One every five  models spanning the \MS\ phase of evolution: from  $X_{\rm c}=0.693$ to
 $X_{\rm c}=0.001$. That means  about 30  models for each set of stellar parameters.
\end{itemize}

An individual  file is provided for each stellar model whose name is formed with
the stellar parameters, for example: m1.00Z0.02X0.70-\#\#\#\#.gong, where ``\#\#\#\#'' can be
``ZAMS'', ``1MS'', ``MS'', ``TAMS'', or a number corresponding to the number of the model.
The format adopted for these files is the {\small{\bf FGONG}}\footnote{Document 
{\it ``Description of the file formats used within ESTA/CoRoT''} at http://www.astro.up.pt/corot/ntools/info.html} 
 one  with the first twenty-five variables. The number of mesh points through the
stellar interior is of the order of  2200, and 100 additional mesh points are used to describe
the stellar atmosphere from an optical depth $\tau=2/3$
outwards, up to  $\tau=10^{-3}$.

\begin{figure}[htbp]
\centering
{\includegraphics[scale=0.4]{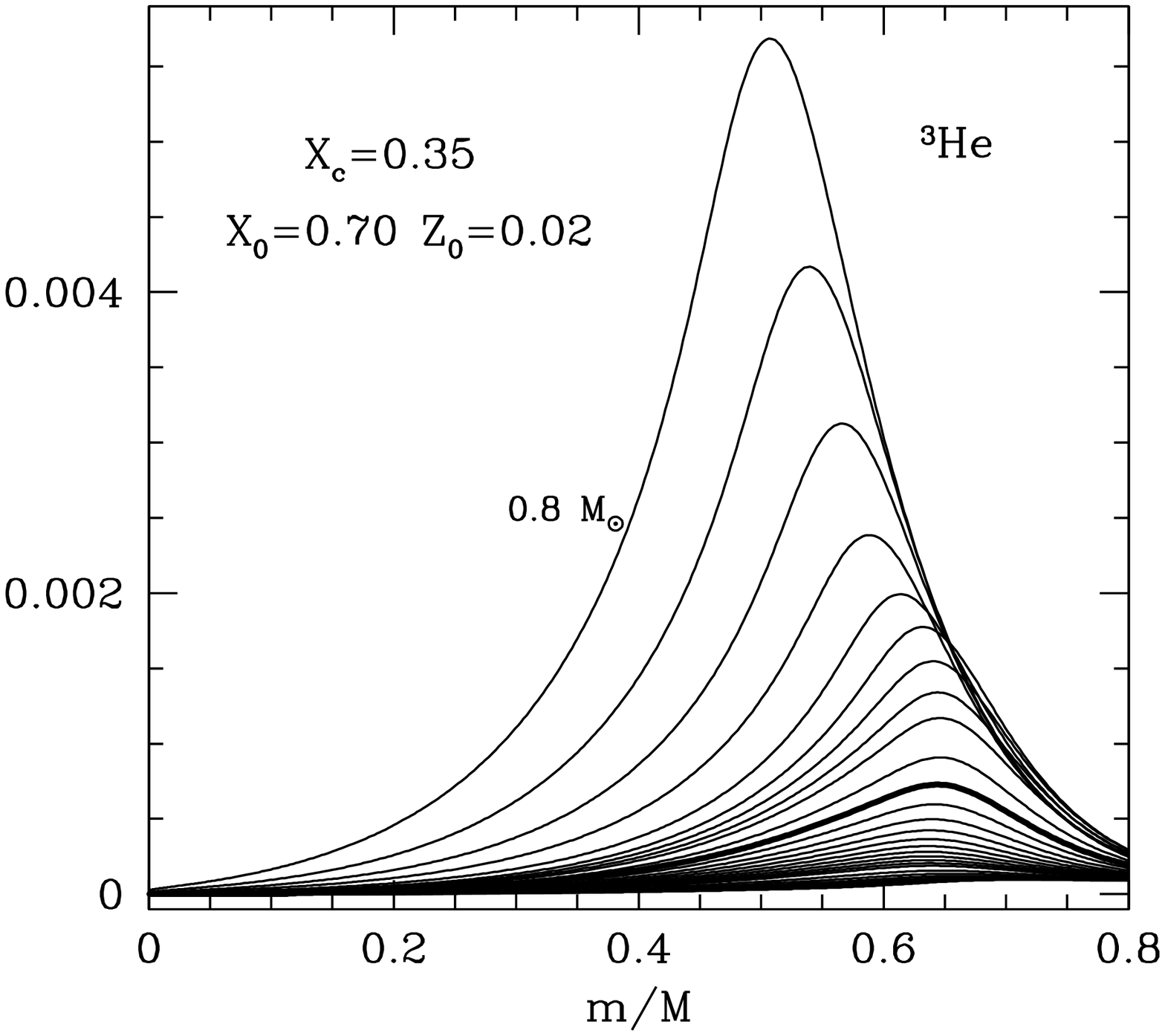}}
{\includegraphics[scale=0.4]{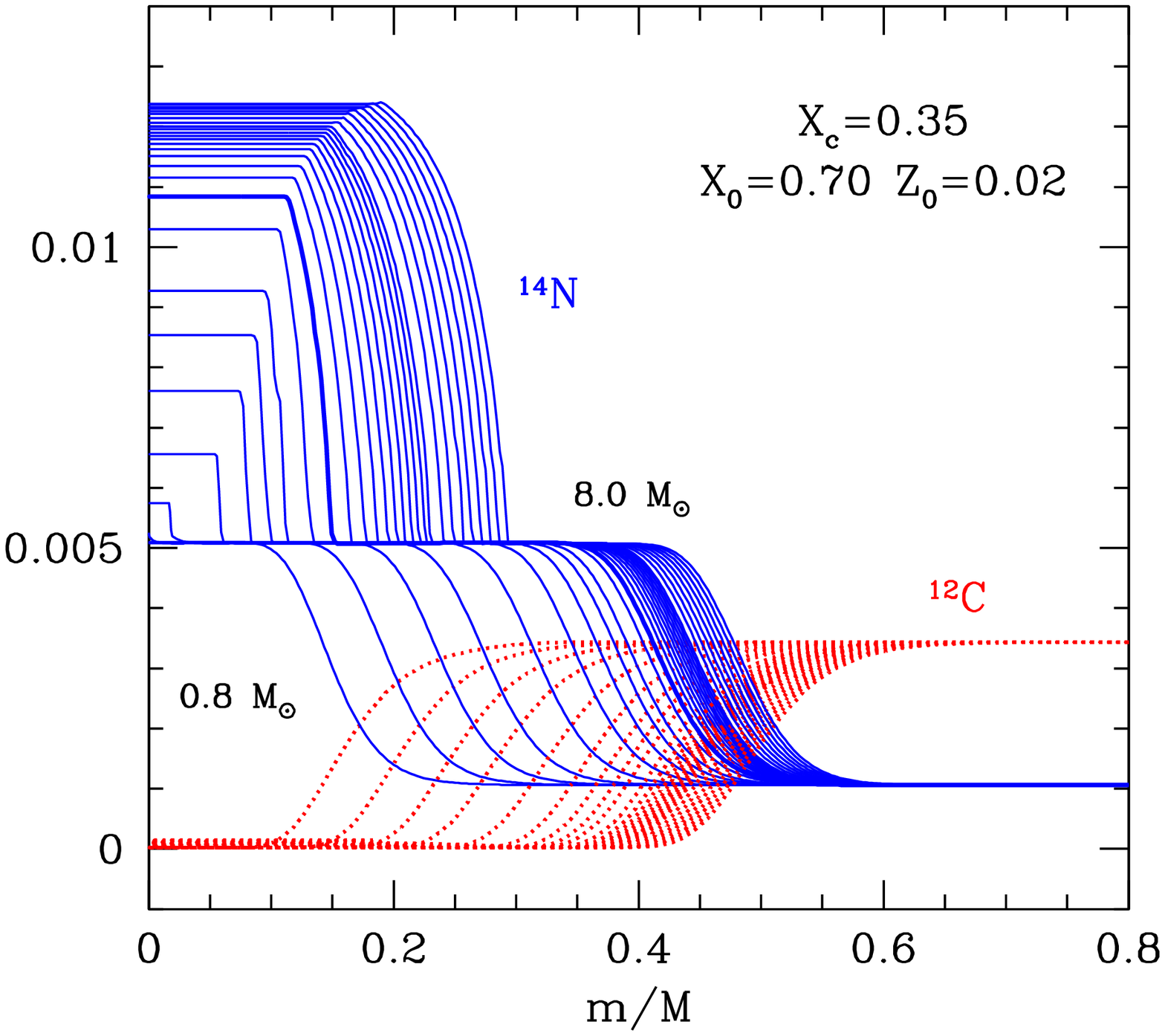}}
\caption{Upper panel: equilibrium abundance of $^3$He for models with masses from
0.8 to 8~\msol. Lower panel: equilibrium abundance of $^{14}$N (solid lines) and
$^{12}$C (dotted lines) for each of the considered stellar masses.
The thicker lines correspond to the chemical profile in a  2~\msol\ model.} 
\label{hecn}
\end{figure}

Among the physical quantities stored in  {\small{\bf FGONG}}\  files there are the mass fraction of the most
important contributors to nuclear energy generation. For illustration, we show in Figs.~\ref{hecn} 
the chemical profile of key elements such as $^3$He, $^{12}$C, and $^{14}$N  
inside the stellar models in the mid--\MS.

\section{Seismic properties}
\label{freqs}
For each equilibrium model in {\small FGONG} format, we also provide  a file containing
the properties of radial and non radial oscillation modes with spherical degree
$\ell=1,$ 2, and 3. These properties  were computed by using the  code \losc\ \citep{rs2-apss} 
with the standard surface boundary condition (regularity of solution when $P=0$ at the surface, that is,
$\delta P/P + (4+\omega^2)\delta r/r=0$).
The angular frequencies of the computed oscillation  modes cover the domain given by
0.3--50 times the dynamical time ($\tau_{\rm dym}=\sqrt{R^3/G\,M}$).

The file name is the same as  that of the equilibrium model but with the extension ``.freq'';
the content is the following: 

\begin{itemize}
\item {\bf Column 1} -- spherical angular degree  $\ell$.
\item {\bf Column 2} -- radial order $n$.
\item {\bf Column 3} -- dimensionless frequency $\omega=\sigma\cdot\tau_{\rm dym}$.
\item {\bf Column 4} -- angular frequency $\sigma$
\item {\bf Column 5} -- frequency $\nu=\sigma/2\,\pi$ in Hz.
\item {\bf Column 6} -- period $P$ in seconds.
\item {\bf Column 7} -- $\beta_{n\ell}=1-C_{n \ell}$ gives the frequency  shift generated by solid rotation, where
$C_{n\ell}$ is the Ledoux's constant \citep{ledoux51}.
\item {\bf Column 8} -- fraction of the kinetic energy associated to the radial component of the motion
                        $ev=E_{\rm kin,V}/E_{\rm kin}$
\item {\bf Column 9} -- $x_{\rm m}=\langle x \rangle$  is  the mean value of $x=r/R$ weighted by
 the kinetic energy of the oscillation mode, and determines in which region the mode is trapped.
\item {\bf Column 10} -- $\Delta=2\sqrt{\langle x^2\rangle-\langle x \rangle^2}$, that takes values between 0 and 1, 
 vanishes for a perfectly trapped mode.
\end{itemize}

Since discontinuities in the sound speed derivatives   in the stellar interior  (such as those
introduced by the boundaries of the convective regions and by the second helium ionization zone)
produce periodic signatures in the frequencies of low degree modes \citep[see e.g.][]{1990LNP...367..283G},
we decided to provide, as well, files  containing the time variations of some quantities that may be
of interest in the seismic analysis of stellar models. So,  for each stellar parameter set $(M,X,Z)$, 
but only for the \MS-lifetime, we built a file whose name  is  formed, as for \HR\ diagram ones,
with the values of the stellar parameters (for instance `` m1.00Z0.02X0.70-ACC.txt''), and whose
content in the columns 7--13 is the following:

\begin {itemize}
\item {\bf Cutoff frequency} (in $\mu$Hz) at the photosphere (column 7):
 $\nu_{\rm ac}=c/(4\pi H_{\rm p})$ where $c$ is the sound speed,  and $H_{\rm p}$ 
the pressure scale height.
\item {\bf Acoustic radius} in seconds (column 8): $r_{\rm ac}=\int_0^R 1/c\,dr$ where $R$ is the 
stellar radius at the photosphere.
\item {\bf Acoustic depth} in seconds of   the border of the convective core (column 10) 
corresponding to the linear relative radius $r_{\rm cc}/R$ given in column 9.
\item {\bf  Acoustic depth} in seconds of the bottom of the convective envelope (column 12) 
corresponding to the linear relative radius
$r_{\rm ce}/R$ given in column 11.
\item {\bf Second He ionization region}: linear relative radius \\($r_{\rm HeII}/R$) in column 12, and the corresponding  acoustic depth $\tau_{\rm HeII}$ in column 13.
\end{itemize}
\noindent
The four first lines, and the   columns 1--6 are the same as in \HRD2-files.

\begin{figure}[htbp]
\centering
{\includegraphics[scale=0.4]{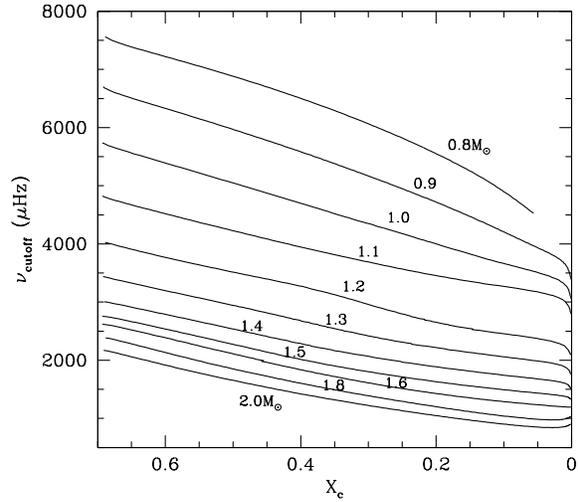}}
\caption{Evolution of the cutoff frequency along the \MS\ evolution of models with masses from
0.8~\msol\ (upper curve) to 2~\msol\ (lower curve).} 
\label{nuacc}
\end{figure}

In principle, only oscillation modes with frequencies lower than the acoustic cutoff one are trapped in the
stellar interior and should be observed. As an indication of the order of magnitude of the expected
oscillation frequencies  we show in Fig.~\ref{nuacc} the variation with time of the cutoff 
frequency for models with masses smaller than 2.~\msol. 

\subsection{Effect of atmospheric boundary conditions}

\begin{figure}[htbp]
\centering
{\includegraphics[scale=0.4]{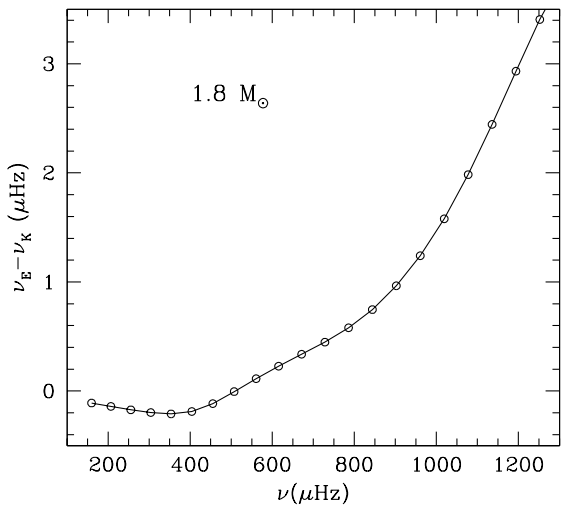}}
\caption{Difference between the oscillation frequencies of 1.8~\msol\ models computed with Eddington's
grey law ($\nu_{\rm E}$) and with Kurucz's atmospheres ($\nu_{\rm K}$).} 
\label{deltaEK}
\end{figure}

For the chemical composition $X=0.70$ and $Z=0.02$ there are two types of stellar models depending
on whether the boundary conditions at $T=T_{\rm eff}$  are given by Kurucz atmosphere models or by
grey ones (with Eddington's law). 
We showed in Fig.~\ref{pms} and \ref{ms} that the  effect on the \HRD\ location was significant
only for $T_{\rm eff} \lessapprox 6300$~K. The effect on the oscillation frequencies may be, however,
 important even for higher temperatures. As an example we show in Fig.~\ref{deltaEK} the difference between 
oscillation  frequencies for 1.8~\msol\ models in  mid-\MS. The difference in  the global parameters for 
these Eddington and Kurucz stellar models are $\Delta R/R=4\times10^{-4}$, $\Delta L/L=1.4\times10^{-5}$, and
$\Delta X_{\rm c} < 10^{-5}$. Nevertheless, the density at $T=T_{\rm eff}$ provided by the two atmosphere
models is different enough to affect a significant fraction of the star, leading to an important
variation of the oscillation frequencies.
Note that for this comparison  only the subphotospheric stellar structure was considered in the
computation of the oscillation frequencies. The differences at high frequency will be even larger if  
the  atmosphere structure  is included  in the calculation.

\section{Conclusions}
\label{conclu}
In order to provide a reference  grid of stellar models  for the first interpretation of the
CoRoT data, we have computed three sets of grey stellar models for 29 different
masses from 0.8 to 8~\msol. Each grid is computed with 
a  different chemical composition, 
and altogether they span metallicity values from [M/H]=-0.45 to +0.07 (taking as 
reference the value of\\ $Z/X$=0.0245 for the Sun given by GN93).
There are $3\times 29\times 3=261$ files containing data on time evolution of global quantities,
around 3050 files with the detailed stellar structure (interior and atmosphere), and
other 3050 with the corresponding oscillation frequencies for low degree modes ($\ell=0$, 1, 2, and 3).
In addition,  a set of stellar models with the Kurucz's atmospheres as boundary conditions
were also computed, for the same values of stellar mass, but for only one chemical composition.
That means additional $\sim$2087 files, 87 for the time evolution, and the rest of them for
the stellar structure and the corresponding seismic properties.

\begin{acknowledgements}
We acknowledge financial support from the Belgian Science Policy Office (BELSPO)
in the frame of the ESA PRO\-DEX~8 program (contract C90199)
 and from the Fonds National de la Recher\-che Scientifique (FNRS).
\end{acknowledgements}


\end{document}